\documentclass[letterpaper,11pt]{article}

\usepackage{times} 
\usepackage{fullpage} 
\usepackage{amsmath} 
\usepackage{amssymb} 
\usepackage{amsthm} 
\usepackage{bbm} 
\usepackage{graphicx} 

\usepackage{tabularx}

\usepackage{ifpdf}
\ifpdf
\usepackage[letterpaper=true,pdftex,bookmarks,
	plainpages=false, 
        pdfpagelabels=true 
        pdftitle=On\ a\ measure\ of\ distance\ for\ quantum\ strategies, 
        pdfauthor=Gus\ Gutoski	 
        ]{hyperref}
\else
\fi

\newcommand{\ot}{\otimes}
\newcommand{\inner}[2]{\langle #1 , #2\rangle}
\newcommand{\Inner}[2]{\left\langle #1 , #2\right\rangle}
\newcommand{\defeq}{\stackrel{\smash{\textnormal{\tiny def}}}{=}}
\newcommand{\kprod}[3]{#1_{#2\dots#3}}

\newcommand{\identity}{\mathbbm{1}}
\newcommand{\idsup}[1]{\identity_{#1}}

\def\Jamiolkowski{J}
\newcommand{\jam}[1]{\Jamiolkowski\pa{#1}}

\DeclareMathOperator{\diag}{diag}

\newtheorem{theorem}{Theorem}
\newtheorem{lemma}[theorem]{Lemma}

\newtheorem{proposition}[theorem]{Proposition}
\newtheorem{fact}[theorem]{Fact}

\theoremstyle{definition}
\newtheorem{defn}[theorem]{Definition}
\newenvironment{definition}{\begin{defn}}{\qed\end{defn}}

\newcommand{\br}[1]{[#1]}

\newcommand{\pa}[1]{(#1)}
\newcommand{\Pa}[1]{\left(#1\right)}
\newcommand{\set}[1]{\{#1\}}
\newcommand{\Set}[1]{\left\{#1\right\}}

\newcommand{\s}[1]{{\br{#1}}}

\DeclareMathOperator{\trace}{Tr}
\newcommand{\ptr}[2]{\trace_{#1}\pa{#2}}
\newcommand{\Ptr}[2]{\trace_{#1}\Pa{#2}}

\newcommand{\tinyspace}{\mspace{1mu}}
\newcommand{\abs}[1]{|\tinyspace#1\tinyspace|}

\newcommand{\norm}[1]{\lVert\tinyspace#1\tinyspace\rVert}
\newcommand{\Norm}[1]{\left\lVert\tinyspace#1\tinyspace\right\rVert}

\newcommand{\tnorm}[1]{\norm{#1}_{\trace}}
\newcommand{\Tnorm}[1]{\Norm{#1}_{\trace}}
\newcommand{\dnorm}[1]{\norm{#1}_{\diamond}}
\newcommand{\Dnorm}[1]{\Norm{#1}_{\diamond}}
\newcommand{\snorm}[2]{\norm{#1}_{\diamond{#2}}}
\newcommand{\Snorm}[2]{\Norm{#1}_{\diamond{#2}}}

\newcommand{\fontmapset}{\mathbf} 

\newcommand{\mset}[2]{\fontmapset{#1}\pa{#2}}

\newcommand{\lin}[1]{\mset{L}{#1}}

\newcommand{\her}[1]{\mset{Her}{#1}}

\newcommand{\pos}[1]{\mset{Pos}{#1}}

\newcommand{\den}[1]{\pos{#1}}

\newcommand{\mea}[1]{\pos{#1}}

\def\cD{\mathcal{D}}

\def\cP{\mathcal{P}}
\def\cQ{\mathcal{Q}}

\def\cW{\mathcal{W}}
\def\cX{\mathcal{X}}
\def\cY{\mathcal{Y}}
\def\cZ{\mathcal{Z}}

\def\bA{\mathbf{A}}
\def\bB{\mathbf{B}}

\def\bS{\mathbf{S}}
\def\bT{\mathbf{T}}

\begin{document}

\title{On a measure of distance for quantum strategies}

\author{
  Gus Gutoski \\[3mm]
  {\small\it
  \begin{tabular}{c}
    Institute for Quantum Computing and School of Computer Science \\
    University of Waterloo,
    Waterloo, Ontario, Canada
  \end{tabular}
  }
}

\date{October 6, 2011\\ \small{(Revised: February 17, 2012)}}

\maketitle

\begin{abstract}

The present paper studies an operator norm that captures the distinguishability of quantum strategies in the same sense that the trace norm captures the distinguishability of quantum states or the diamond norm captures the distinguishability of quantum channels.
Characterizations of its unit ball and dual norm are established via strong duality of a semidefinite optimization problem.
A full, formal proof of strong duality is presented for the semidefinite optimization problem in question.
This norm and its properties are employed to generalize a state discrimination result of Ref.~\cite{GutoskiW05}.
The generalized result states that for any two convex sets $\bS_0,\bS_1$ of strategies there exists a fixed interactive measurement scheme that successfully distinguishes any choice of $S_0\in\bS_0$ from any choice of $S_1\in\bS_1$ with bias proportional to the minimal distance between the sets $\bS_0$ and $\bS_1$ as measured by this norm.
A similar discrimination result for channels then follows as a special case.

\end{abstract}

\section{Introduction}

\subsection{Quantum strategies}

A \emph{quantum strategy} is a complete specification of the actions of one party in an interaction involving the exchange of multiple rounds of quantum messages with one or more other parties.
Fundamental objects in the study of quantum information such as states, measurements, and channels may be viewed as special cases of strategies.
A particularly useful representation for quantum strategies is presented in Ref.~\cite{GutoskiW07}.
(See also Ref.~\cite{ChiribellaD+09a}.)



Briefly and informally, this representation associates with each strategy a single positive semidefinite operator $S$, the dimensions of which depend upon the size of the messages exchanged in the interaction.
It is shown in Ref.~\cite{GutoskiW07} that the set of all positive semidefinite operators which are valid representations of strategies is characterized by a simple and efficiently-verifiable collection of linear equality conditions.
(Essentially, these conditions reflect the intuitive causality constraint that outgoing messages in early rounds of the interaction cannot depend upon incoming messages from later rounds.)
An explicit list of these conditions is given in Section \ref{sec:review}.

In order to extract useful classical information from an interaction, a strategy might call for one or more quantum measurements throughout the interaction.
In this case, the strategy is instead represented by a set $\set{S_a}$ of positive semidefinite operators indexed by all the possible combinations of outcomes of the measurements.
These strategies are called \emph{measuring strategies} and satisfy $\sum_a S_a = S$ for some ordinary (non-measuring) strategy $S$.
(By comparison, an ordinary POVM-type quantum measurement $\set{P_a}$ satisfies $\sum_a P_a = I$.)

Conveniently, the relationship between measuring and non-measuring strategies is analogous to that between ordinary measurements and states.
In particular, the postulates of quantum mechanics dictate that for any ordinary measurement $\set{P_a}$ with outcomes indexed by $a$ and any density operator $\rho$ it holds that the probability with which $\set{P_a}$ yields outcome $a$ when applied to a quantum system whose state is represented by $\rho$ is given by the inner product
\[ \Pr[\textrm{$\set{P_a}$ yields outcome $a$ on $\rho$}] = \inner{P_a}{\rho} = \ptr{}{P_a\rho}.\]
Similarly, it is shown in Ref.~\cite{GutoskiW07} that the probability with which a measuring strategy $\set{S_a}$ yields outcome $a$ after an interaction with a compatible quantum strategy $T$ is given by
\[ \Pr[\textrm{$\set{S_a}$ yields outcome $a$ when interacting with $T$}] = \inner{S_a}{T} = \ptr{}{S_aT}.\]
A more formal review of quantum strategies is given in Section \ref{sec:review}.

\subsection{Distance measures}

In the study of quantum information the need often arises for a distance measure that quantifies the observable difference between two states or channels.
For states, such a distance measure is induced by the {trace norm} 
of the difference between two density operators.
For channels, the measure of choice is induced by the {diamond norm} 
of the difference between two completely positive and trace-preserving linear maps.

The use of the trace norm to measure distance between states can be traced back to the 1960s (see Nielsen and Chuang \cite[Chapter 9]{NielsenC00}).
The diamond norm was defined by Kitaev for the explicit purpose of measuring distance between channels \cite{Kitaev97,AharonovK+98}.
It was later noticed that the diamond norm is related via the notion of duality to the \emph{norm of complete boundedness} for linear maps, an object of study in mathematics circles since the 1980s.
(See Paulsen \cite{Paulsen02}.)

A suitable distance measure for quantum strategies was first considered by Chiribella, D'Ariano, and Perinotti \cite{ChiribellaD+08b}.
This distance measure captures the distinguishability of quantum strategies in the same sense that the trace norm captures the distinguishability of states or the diamond norm captures the distinguishability of channels.
Given the strikingly similar relationships between states and measurements and between strategies and measuring strategies, the new distance measure suggests itself:
whereas the trace norm $\tnorm{\rho-\sigma}$ for quantum states $\rho,\sigma$ is easily seen to satisfy
\[ \tnorm{\rho-\sigma} = \max \Set{\inner{P_0-P_1}{\rho-\sigma} : \set{P_0,P_1} \textrm{ is a quantum measurement} }, \]
the \emph{strategy $r$-norm} $\snorm{R-S}{r}$ for quantum strategies $R,S$ can be informally defined by
\[ \snorm{R-S}{r} \defeq \max \Set{\inner{T_0-T_1}{R-S} : \set{T_0,T_1} \textrm{ is a compatible measuring strategy} }.\]
(A formal definition of this norm appears in Section \ref{sec:new-norm} after due discussion of preliminary material.)

Here the subscript $r$ denotes the number of rounds of messages in the protocol for which $R,S$ are strategies.
In particular, each positive integer $r$ induces a different strategy norm.
The choice of notation is inspired by the fact that this norm coincides with the diamond norm for the case $r=1$ \cite{ChiribellaD+08b}.
Hence, the strategy $r$-norm can be viewed as a generalization of the diamond norm for Hermitian-preserving linear maps.

Little else is known of the strategy $r$-norm.
It was noted in Ref.~\cite{ChiribellaD+08b} that this norm differs from the diamond norm for $r>1$.
The norm was also mentioned in Refs.~\cite{ChiribellaD+08d,ChiribellaD+09a} and Chiribella \emph{et al.}~proved a continuity bound for the strategy $r$-norm as part of their short impossibility proof for quantum bit commitment schemes \cite{ChiribellaD+09b}.

Recent work in the mathematical physics literature has focussed on an extension of the diamond norm (or, equivalently, the norm of complete boundedness) to $k$-minimal and $k$-maximal operator spaces and operator systems and the relationships of these norms and spaces with entangled quantum states and with the $k$-positive and $k$-superpositive cones of operators---see Refs.\ \cite{JohnstonK+11,SkowronekS+09} and the references therein.
We briefly elaborate upon this extension of the diamond norm at the end of Section \ref{sec:new-norm:diamond-norm}.
However, all appearances indicate that the these objects have little to do with the strategy $r$-norm or the cone generated by $r$-round strategies.

\subsection{Results}

Characterizations of the unit balls of the strategy $r$-norm and its dual norm are presented as Theorem \ref{thm:unit-ball} in Section \ref{sec:unit-ball-duality}.
This theorem is proven via strong duality of a semidefinite optimization problem.
A full, formal proof of strong duality for this problem is given in Appendix \ref{appendix:duality}.

These characterizations are then used to generalize a state discrimination result of Ref.~\cite{GutoskiW05}, which asserts that for any two convex sets $\bA_0,\bA_1$ of states there exists a fixed measurement that successfully distinguishes any choice of $\rho_0\in\bA_0$ from any choice of $\rho_1\in\bA_1$ with bias proportional to the minimal trace norm distance between the sets $\bA_0$ and $\bA_1$.

By analogy,
Theorem \ref{thm:sep}
in Section \ref{sec:distinguish-convex}
asserts that for any two convex sets $\bS_0,\bS_1$ of $r$-round strategies there exists a fixed compatible $r$-round measuring strategy that successfully distinguishes any choice of $S_0\in\bS_0$ from any choice of $S_1\in\bS_1$ with bias proportional to the minimum distance between the sets $\bS_0$ and $\bS_1$ as measured by the strategy $r$-norm.
Just as in Ref.~\cite{GutoskiW05}, it therefore follows that
\begin{enumerate}

\item
This compatible measuring strategy can be used to discriminate between \emph{any} choices of strategies from $\bS_0,\bS_1$ at least as well as \emph{any other} compatible measuring strategy could discriminate between the two \emph{closest} strategies from those sets.

\item
Even if two (or more) distinct pairs $(S_0,S_1)$ and $(S_0',S_1')$ both minimize the distance between $\bS_0$ and $\bS_1$ then \emph{both} pairs may be \emph{optimally} discriminated by the \emph{same compatible measuring strategy}.

\end{enumerate}
As a special case of Theorem \ref{thm:sep}, a similar discrimination result is obtained for convex sets of channels with the diamond norm in place of the strategy $r$-norm.


Strong duality of the aforementioned semidefinite optimization problem also yields an alternate and arguably simpler proof of a property of strategies established in Ref.~\cite{GutoskiW07}.
This property, listed as Theorem \ref{thm:max-output-prob}
in Section \ref{sec:unit-ball-duality:max-output}, establishes a useful formula for the maximum probability with which a measuring strategy can be forced to produce a given measurement outcome by a compatible interacting strategy.

\subsection{Notation}

The following table summarizes the notation used in this paper.

\begin{center}
\begin{tabularx}{\textwidth}{lX}
$\cW,\cX,\cY,\cZ$ & Calligraphic letters denote finite-dimensional complex Euclidean spaces of the form $\mathbb{C}^n$.\\
$\kprod{\cX}{1}{n}$ & Shorthand notation for the tensor product $\cX_1\ot\cdots\ot\cX_n$.\\
$\bS,\bT,\bA,\bB$ & Bold letters denote sets of operators.\\
$\lin{\cX}$ & The (complex) space of all linear operators $A:\cX\to\cX$, implicitly identified with $\mathbb{C}^{n\times n}$.\\
$\her{\cX}$ & The (real) subspace of Hermitian operators within $\lin{\cX}$.\\
$\pos{\cX}$ & The cone of positive semidefinite operators within $\her{\cX}$.\\
$\succeq,\succ,\preceq,\prec$ & The semidefinite partial ordering on $\her{\cX}$.\\
$A^*$ & The adjoint of an operator $A:\cX\to\cY$, which has the form $A^*:\cY\to\cX$.\\
$\inner{A}{B}$ & The standard inner product between two operators $A,B:\cX\to\cY$.  Defined by $\inner{A}{B}\defeq\ptr{}{A^*B}$.\\
$I_\cX$ & The identity operator acting on $\cX$.\\
$\idsup{\cX}$ & The identity linear map acting on $\lin{\cX}$.\\
$\trace_\cX$ & The partial trace over $\cX$.  For any space $\cY$ this linear map is defined by
\[ \trace_\cX:\lin{\cX\ot\cY}\to\lin{\cY}:X\ot Y\mapsto \ptr{}{X} Y. \]
(This definition extends to all of $\lin{\cX\ot\cY}$ by linearity on operators of the form $X\ot Y$.)\\
$\jam{\Phi}$ & The Choi-Jamio\l kowski operator representation of a linear map $\Phi$.  (See below.)
\end{tabularx}
\end{center}

A \emph{density operator} or \emph{quantum state} is a positive semidefinite operator with unit trace.
A \emph{quantum measurement} with (finitely many) outcomes indexed by $a$ is a finite set $\set{P_a}\subset\pos{\cX}$ of positive semidefinite operators with $\sum_a P_a = I_\cX$.

A linear map $\Phi$ is \emph{positive} if $\Phi(X)\succeq 0$ whenever $X\succeq 0$ and \emph{completely positive} if $\Phi\ot\idsup{\cW}$ is positive for all choices of the space $\cW$.
A linear map $\Phi$ is \emph{trace-preserving} if $\ptr{}{\Phi(X)}=\ptr{}{X}$ for all $X$.
As usual, the set of all possible physically realizable operations on quantum states is identified with the set of completely positive and trace-preserving linear maps.
Such a map is often called a \emph{channel}.

The \emph{Choi-Jamio\l kowski isomorphism} associates with each linear map $\Phi:\lin{\cX}\to\lin{\cY}$ a unique operator $\jam{\Phi}\in\lin{\cY\ot\cX}$ via the formula \[ \jam{\Phi} = \sum_{i,j=1}^{\dim(\cX)} \Phi(E_{i,j})\ot E_{i,j} \]
where $\set{E_{i,j}}$ is the standard orthonormal basis for $\lin{\cX}$.
It holds that $\Phi$ is completely positive if and only if $\jam{\Phi}$ is positive semidefinite and that that $\Phi:\lin{\cX}\to\lin{\cY}$ is trace-preserving if and only if $\ptr{\cY}{\jam{\Phi}}=I_\cX$.
A linear map $\Phi$ is \emph{Hermitian-preserving} if $\Phi(X)$ is Hermitian whenever $X$ is Hermitian.
It holds that $\Phi$ is Hermitian-preserving if and only if $\jam{\Phi}$ is a Hermitian operator.

\subsection{Table of contents}

The rest of this paper is organized as follows.\\

\begin{tabular}{rl}
Section \ref{sec:review} & Review of quantum strategies \\
Section \ref{sec:new-norm} & Discrimination problems and norms \\
Section \ref{sec:unit-ball-duality} & Unit ball of the strategy $r$-norm and its dual \\
Section \ref{sec:distinguish-convex} & Distinguishability of convex sets of strategies \\
Appendix \ref{appendix:duality} & Appendix to Section \ref{sec:unit-ball-duality}: formal proof of semidefinite optimization duality
\end{tabular}

\section{Review of quantum strategies}
\label{sec:review}

This section reviews the formalism of quantum strategies as presented in Ref.~\cite{GutoskiW07}.
The curious reader is referred to Refs.~\cite{GutoskiW07,ChiribellaD+09a} for additional detail.

\subsection{Operational formalism}
\label{sec:review:operational}

At a high level, a \emph{strategy} is a complete description of one party's actions in a multiple-round interaction involving the exchange of quantum information with one or more other parties.
For convenience, let us call this party \emph{Alice}.
As we are only concerned for the moment with Alice's actions during the interaction, it is convenient to bundle the remaining parties into one party, whom we call \emph{Bob}.

From Alice's point of view every finite interaction decomposes naturally into a finite number $r$ of \emph{rounds}.
In a typical round a message comes in, the message is processed, and a reply is sent out.
Naturally, this reply might depend upon messages exchanged during previous rounds of the interaction.
To account for such a dependence, we allow for a memory workspace to be maintained between rounds.

The complex Euclidean spaces corresponding to the incoming and outgoing messages in an arbitrary round $i$ shall be denoted $\cX_i$ and $\cY_i$, respectively.
The space corresponding to the memory workspace to be stored for the next round shall be denoted $\cZ_i$.
In a typical round $i$ of the quantum interaction, Alice's actions are faithfully represented by a channel
\[ \Phi_i : \lin{\cX_i\ot\cZ_{i-1}}\to\lin{\cY_i\ot\cZ_i}. \]
The first round of the interaction is a special case: there is no need for an incoming memory space for this round, so the channel $\Phi_1$ has the form
\[ \Phi_1 : \lin{\cX_1}\to\lin{\cY_1\ot\cZ_1}. \]
The final round of the interaction is also a special case: there is no immediate need for an outgoing memory space for this round.
However, the presence of this final memory space better facilitates the forthcoming discussion of strategies involving measurements.
Thus, the channel $\Phi_r$ representing Alice's actions in the final round of the interaction has the same form as those from previous rounds:
\[ \Phi_r : \lin{\cX_r\ot\cZ_{r-1}}\to\lin{\cY_r\ot\cZ_r}. \]


In order to extract classical information from the interaction it suffices to permit Alice to perform a single quantum measurement on her final memory workspace.
(Sufficiency of a single measurement at the end of the interaction follows immediately from foundational results on mixed state quantum computations \cite{AharonovK+98}, which tell us that any process calling for one or more intermediate measurements can be efficiently simulated by a channel with a single measurement at the end.)

Formally then, the \emph{operational description} of an \emph{$r$-round strategy} for an interaction with \emph{input spaces} $\cX_1,\ldots,\cX_r$ and \emph{output spaces} $\cY_1,\ldots,\cY_r$ is specified by:
\begin{enumerate}

\item[1.]
  Complex Euclidean spaces $\cZ_1,\ldots,\cZ_r$, called \emph{memory spaces}, and

\item[2.]
  An $r$-tuple of channels $(\Phi_1,\ldots,\Phi_r)$ of the form
  \begin{align*}
    \Phi_1 &: \lin{\cX_1}\to\lin{\cY_1\ot\cZ_1}\\
    \Phi_i &: \lin{\cX_i\ot\cZ_{i-1}}\to\lin{\cY_i\ot\cZ_i} \quad (2 \leq i \leq r).
  \end{align*}

\end{enumerate}
The operational description of an $r$-round \emph{measuring} strategy with outcomes indexed by $a$ is specified by items 1 and 2 above, as well as:
\begin{enumerate}

\item[3.]
  A measurement $\set{P_a}\subset\mea{\cZ_r}$ on the last memory space $\cZ_r$.

\end{enumerate}
We use the words ``operational description'' to distinguish this representation for strategies from the representation to be described in Section \ref{sec:review:choijam}.

\begin{figure}
  \begin{center}
\includegraphics{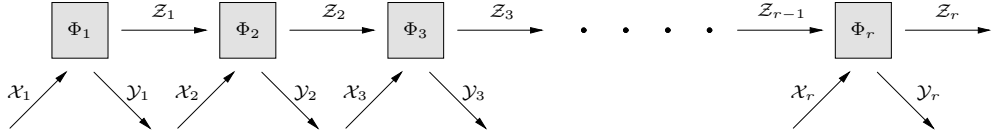}
  \end{center}
  \caption{An $r$-round strategy.}
  \label{fig:strategy}
\end{figure}

A strategy without a measurement is referred to a \emph{non-measuring} strategy.
A non-measuring strategy may be viewed as a measuring strategy in which the measurement has only one outcome, so that $\set{P_a}=\set{I}$ is the singleton set containing the identity.
Figure~\ref{fig:strategy} illustrates an $r$-round non-measuring strategy.

Note that input and output spaces may have dimension one, which corresponds to an empty message.
One can therefore view simple actions such as the preparation of a quantum state or performing a measurement without producing a quantum output as special cases of strategies.
(Special cases such as this are also discussed at the end of Section \ref{sec:review:properties}.)

\begin{figure}
  \begin{center}
\includegraphics{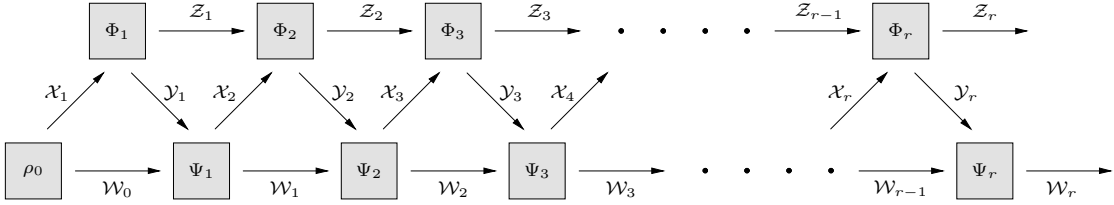}
  \end{center}
  \caption{An interaction between an $r$-round strategy and co-strategy.}
  \label{fig:interaction}
\end{figure}

In order for interaction to occur Bob must supply the incoming messages $\cX_1,\dots,\cX_r$ and process the outgoing messages $\cY_1,\dots,\cY_r$ as suggested by Figure \ref{fig:interaction}.
Due to the inherently asymmetric nature of any interaction (only one of the parties can send the first message or receive the final message), the actions of Bob are described not by a \emph{strategy}, but by a slightly different object called a \emph{co-strategy}.

Formally, the operational description of an $r$-round \emph{co-strategy} for an interaction with input spaces $\cX_1,\ldots,\cX_r$ and output spaces $\cY_1,\ldots,\cY_r$ is specified by:
\begin{enumerate}

\item[1.]
  Complex Euclidean memory spaces $\cW_0,\ldots,\cW_r$,

\item[2.]
  A quantum state $\rho_0 \in \den{\cX_1\otimes\cW_0}$, and

\item[3.]
  An $r$-tuple of channels $(\Psi_1,\ldots,\Psi_r)$ of the form
  \begin{align*}
    \Psi_i & : \lin{\cY_i\ot\cW_{i-1}}\to\lin{\cX_{i+1}\ot\cW_i} \quad (1\leq i\leq r-1)\\
    \Psi_r & : \lin{\cY_r\ot\cW_{r-1}}\to\lin{\cW_r}.
  \end{align*}

\end{enumerate}
The operational description of an $r$-round \emph{measuring} co-strategy with outcomes indexed by $b$ is specified by items 1, 2 and 3 above, as well as:
\begin{enumerate}

\item[4.]
  A measurement $\set{Q_b}\subset\mea{\cW_r}$ on the last memory space $\cW_r$.

\end{enumerate}

The \emph{output} of an interaction between a strategy and a co-strategy is the result of the measurements performed after the interaction.
In particular, the postulates of quantum mechanics tell us that the probability with which Alice and Bob output the pair $(a,b)$ is given by
\[ \Pr[\textrm{output $(a,b)$}] = \Ptr{}{\Pa{P_a\ot Q_b}\sigma_r} \]
where $\sigma_r\in\den{\cZ_r\ot\cW_r}$ is the state of the system at the end of the interaction.
(This state is most conveniently described by the recursive formula
\(
  \sigma_{i+1} =
  \pa{\idsup{\cZ_i}\ot\Psi_i}\circ\pa{\Phi_i\ot\idsup{\cW_{i-1}}}(\sigma_i)
\)
with $\sigma_0=\rho_0$.)

\subsection{Choi-Jamio{\l}kowski formalism}
\label{sec:review:choijam}

While intuitive from an operational perspective, the operational description of a strategy by an $r$-tuple of channels and a measurement is often inconvenient.
In this subsection we describe the alternate formalism for strategies presented in Ref.~\cite{GutoskiW07} derived from the Choi-Jamio{\l}kowski representation for linear maps.

\begin{figure}
  \begin{center}
\includegraphics{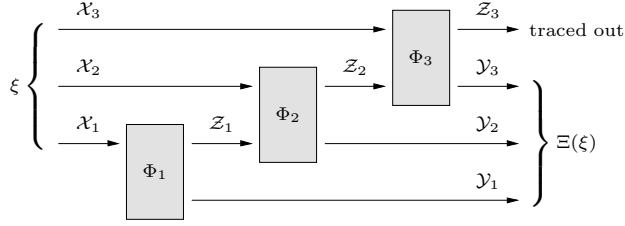}
  \end{center}
  \caption{The linear map $\Xi$ associated with a three-round strategy.}
  \label{fig:choijam}
\end{figure}

Let us first restrict attention to $r$-round non-measuring strategies.
To the $r$-round strategy specified by channels $(\Phi_1,\ldots,\Phi_r)$ we associate a single channel
\[\Xi:\lin{\kprod{\cX}{1}{r}} \rightarrow \lin{\kprod{\cY}{1}{r}}. \]
This channel takes a given $r$-partite input state $\xi\in\den{\kprod{\cX}{1}{r}}$ and feeds the portions of this state corresponding to the input spaces $\cX_1,\ldots,\cX_r$ into the network pictured in Figure~\ref{fig:strategy}, one piece at a time.
The final memory space $\cZ_r$ is then traced out, leaving some element $\Xi(\xi)\in \den{\kprod{\cY}{1}{r}}$.
Such a map is depicted in Figure~\ref{fig:choijam} for the case $r=3$.
An \emph{$r$-round non-measuring strategy} for input spaces $\cX_1,\dots,\cX_r$ and output spaces $\cY_1,\dots,\cY_r$ is defined to be Choi-Jamio{\l}kowski representation
\[\jam{\Xi}\in\pos{\kprod{\cY}{1}{r}\ot\kprod{\cX}{1}{r}}\]
of the channel $\Xi$ we have just described.
(This definition of a strategy is distinguished from the operational description of Section \ref{sec:review:operational} by the absence of the words ``operational description.'')

To a measuring strategy with measurement $\set{P_a}\subset\mea{\cZ_r}$ we associate not a single channel, but instead a set $\set{\Xi_a}$ of linear maps, one for each measurement outcome $a$, each of the same form
\[\Xi_a:\lin{\kprod{\cX}{1}{r}} \rightarrow \lin{\kprod{\cY}{1}{r}}. \]
Each $\Xi_a$ is defined precisely as in the non-measuring case except that the partial trace over $\cZ_r$ is replaced by the mapping
\[ X \mapsto \ptr{\cZ_r}{\pa{P_a\ot I_{\kprod{\cY}{1}{r}}}X}. \]
Each of the linear maps $\Xi_a$ is completely positive and trace non-increasing, but not necessarily trace-preserving.
Notice that \[ \sum_a \Xi_a = \Xi \] where $\Xi$ is the channel defined as in the non-measuring case.
This observation is consistent with the view that $\Xi$ represents a measuring strategy with only one outcome.

\begin{figure}
  \begin{center}
\includegraphics{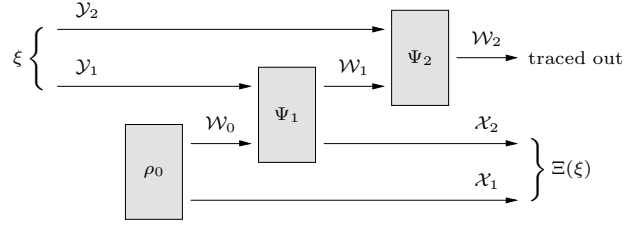}
  \end{center}
  \caption{The linear map $\Xi$ associated with a two-round co-strategy.}
  \label{fig:choijam-co}
\end{figure}

Non-measuring co-strategies are defined similarly to non-measuring strategies except that we take the Choi-Jamio{\l}kowski representation $\jam{\Xi^*}$ of the adjoint linear mapping $\Xi^*$ of the channel $\Xi$ described above.
So, for example, an $r$-round non-measuring co-strategy specified by $(\rho_0,\Psi_1,\dots,\Psi_r)$ induces a channel \[ \Xi:\lin{\kprod{\cY}{1}{r}}\to\lin{\kprod{\cX}{1}{r}} \] as suggested by Figure \ref{fig:choijam-co}.
Notice that the domain $\lin{\kprod{\cY}{1}{r}}$ and range $\lin{\kprod{\cX}{1}{r}}$ are switched when the mapping $\Xi$ is derived from a co-strategy instead of a strategy.
The domain and range are switched back again by working with the adjoint mapping $\Xi^*$.
One implication of this choice to work with the adjoint mapping for co-strategies is that the Choi-Jamio{\l}kowski representations for both strategies and co-strategies are always elements of $\pos{\kprod{\cY}{1}{r}\ot\kprod{\cX}{1}{r}}$.
(Otherwise, co-strategies would lie in $\pos{\kprod{\cX}{1}{r}\ot\kprod{\cY}{1}{r}}$.)

The extension from non-measuring co-strategies to measuring co-strategies is completely analogous to that for strategies.

\subsection{Properties of strategies}
\label{sec:review:properties}

This subsection lists several useful properties of strategies, each of which was first established in Ref.\ \cite{GutoskiW07}.

The first such property is that the set of all linear maps that represent legal non-measuring strategies is conveniently characterized by a collection of linear constraints on the Choi-Jamio\l kowski matrix.
Specifically, an arbitrary operator $S\in\lin{\kprod{\cY}{1}{r}\ot\kprod{\cX}{1}{r}}$ is the representation of some $r$-round non-measuring strategy for input spaces $\cX_1,\dots,\cX_r$ and output spaces $\cY_1,\dots,\cY_r$
if and only if $S$ is positive semidefinite and there exist positive semidefinite operators $S_\s{1},\dots,S_\s{r}$ of the form
\[
  \qquad\qquad\qquad\qquad\qquad
  S_\s{i} \in \pos{\kprod{\cY}{1}{i}\ot\kprod{\cX}{1}{i}}
  \qquad \qquad \qquad(1\leq i\leq r)
\]
such that $S=S_\s{r}$ and
\begin{align*}
\Ptr{\cY_r}{S_\s{r}} &= S_\s{r-1} \ot I_{\cX_r} \\
&\vdots \\
\Ptr{\cY_2}{S_\s{2}} &= S_\s{1} \ot I_{\cX_2} \\
\Ptr{\cY_1}{S_\s{1}} &= I_{\cX_1}.
\end{align*}
In other words, there exist memory spaces $\cZ_1,\dots,\cZ_r$ and channels $(\Phi_1,\dots,\Phi_r)$ such that the channel $\Xi$ induced by these objects as described in Section \ref{sec:review:choijam} satisfies $\jam{\Xi}=S$ if and only if $S$ meets the above criteria.

%
%
Similarly, an operator $T\in\lin{\kprod{\cY}{1}{r}\ot\kprod{\cX}{1}{r}}$ is the representation of some $r$-round non-measuring co-strategy for input spaces $\cX_1,\dots,\cX_r$ and output spaces $\cY_1,\dots,\cY_r$ if and only if $T$ is positive semidefinite and there exist positive semidefinite operators $T_\s{1},\dots,T_\s{r}$ of the form
\[
  \qquad\qquad\qquad\qquad\qquad
  T_\s{i} \in \pos{\kprod{\cY}{1}{i-1}\ot\kprod{\cX}{1}{i}}
  \qquad \qquad \qquad(1\leq i\leq r)
\]
such that
\begin{align*}
T &= T_\s{r} \ot I_{\cY_r} \\
\Ptr{\cX_r}{T_\s{r}} &= T_\s{r-1} \ot I_{\cY_{r-1}} \\
&\vdots \\
\Ptr{\cX_2}{T_\s{2}} &= T_\s{1} \ot I_{\cY_1} \\
\Ptr{}{T_\s{1}} &= 1.
\end{align*}

Measuring strategies also admit a simple characterization: a set $\set{S_a}\subset\pos{\kprod{\cY}{1}{r}\ot\kprod{\cX}{1}{r}}$ is the representation of some $r$-round measuring strategy for input spaces $\cX_1,\dots,\cX_r$ and output spaces $\cY_1,\dots,\cY_r$ if and only if \( \sum_a S_a \) is the representation of some $r$-round non-measuring strategy for the same input and output spaces.
A similar characterization holds for measuring co-strategies.

When an $r$-round measuring strategy $\set{S_a}$ for input spaces $\cX_1,\dots,\cX_r$ and output spaces $\cY_1,\dots,\cY_r$ interacts with an $r$-round measuring co-strategy $\set{T_b}$ for the same input and output spaces the probability with which the output pair $(a,b)$ occurs is given by the inner product
\[ \Pr[\textrm{interaction between $\set{S_a}$ and $\set{T_b}$ yields output $(a,b)$}] = \inner{S_a}{T_b} = \ptr{}{S_aT_b}.\]
The standard inner product relationship between ordinary states and measurements is recovered in the special case $r=1$ and $\dim(\cY_1)=1$.
To see this, notice that the set of all non-measuring co-strategies coincides in this case with the set of all density operators on $\cX_1$.
Any measuring strategy $\set{S_a}\subset\pos{\cX_1}$ satisfies $\sum_a S_a = I_{\cX_1}$ and hence acts as an ordinary measurement on $\cX_1$.
The previous inner product formula therefore tells us
\[ \Pr[\textrm{$\set{S_a}$ yields output $a$ when applied to $\rho$}] = \inner{S_a}{\rho} = \ptr{}{S_a\rho},\]
which is the familiar postulate of quantum mechanics.


\section{Discrimination problems and norms}
\label{sec:new-norm}

A formal definition of the strategy $r$-norm is given in Definition \ref{def:new-norm} in Section \ref{sec:new-norm:strategy-norm} after due discussions of the trace and diamond norms in Sections \ref{sec:new-norm:trace-norm} and \ref{sec:new-norm:diamond-norm}, respectively.
A discrimination problem for convex sets of states, channels, and strategies is discussed in Section \ref{sec:new-norm:discrimination-problems}.

\subsection{The trace norm as a distance measure for states}
\label{sec:new-norm:trace-norm}

The \emph{trace norm} $\tnorm{X}$ of an arbitrary operator $X$ is defined as the sum of the singular values of $X$.
If $X$ is Hermitian then it is a simple exercise to verify that its trace norm is given by
\begin{align*}
\tnorm{X}
&= \max \Set{ \inner{P_0-P_1}{X} \::\: P_0,P_1\succeq 0,\ P_0+P_1=I } \\
&= \max \Set{ \inner{P_0-P_1}{X} \::\: \set{P_0,P_1} \textrm{ is a two-outcome measurement} }.
\end{align*}

The trace norm provides a physically meaningful distance measure for quantum states in the sense that it captures the maximum likelihood with which two states can be correctly discriminated.
This fact is illustrated by a simple example involving two parties called \emph{Alice} and \emph{Bob} and a fixed pair of quantum states $\rho_0,\rho_1$.
Suppose Bob selects a state $\rho\in\set{\rho_0,\rho_1}$ uniformly at random and gives Alice a quantum system prepared in state $\rho$.
Alice has a complete description of both $\rho_0$ and $\rho_1$, but she does not know which of the two was selected by Bob.
Her goal is to correctly guess which of $\set{\rho_0,\rho_1}$ was selected based upon the outcome of a measurement she conducts on $\rho$.

Since Alice's guess is binary-valued and completely determined by her measurement, that measurement can be assumed to be a two-outcome measurement $\set{P_0,P_1}$ wherein outcome $a\in\set{0,1}$ indicates a guess that Bob prepared $\rho=\rho_a$.
The probability with which Alice successfully discriminates $\rho_0$ from $\rho_1$ is easily shown to be
\[
  \Pr[\textrm{Alice guesses correctly}] = 
  \frac{1}{2} + \frac{1}{4} \inner{P_0-P_1}{\rho_0-\rho_1} \leq
  \frac{1}{2} + \frac{1}{4} \tnorm{\rho_0-\rho_1}
\]
with equality achieved at the optimal measurement $\set{P_0,P_1}$ for Alice.
This fundamental observation was originally made by Helstrom \cite{Helstrom69}.

\subsection{The diamond norm as a distance measure for channels}
\label{sec:new-norm:diamond-norm}


The \emph{linear map trace norm} is induced by the operator trace norm via the formula \[ \tnorm{\Phi} \defeq \max_{\tnorm{X}=1} \tnorm{\Phi(X)}. \]
Unfortunately, this norm does not lead to an overly useful distance measure for quantum channels.
To achieve such a measure, the trace norm must be ``stabilized'' to form the \emph{diamond norm} via the formula
\[ \dnorm{\Phi} \defeq \sup_\cW \Tnorm{\Phi\ot\idsup{\cW}} \]
where the supremum is taken over all finite-dimensional complex Euclidean spaces $\cW$.

Much is known of the diamond norm.
For example, if $\Phi$ has the form $\Phi:\lin{\cX}\to\lin{\cY}$ then the supremum in the definition of $\dnorm{\Phi}$ is always achieved by some space $\cW$ whose dimension does not exceed that of the input space $\cX$.
(This fact was originally established for the completetly bounded norm by Smith \cite{Smith83} and independently rediscovered for the diamond norm by Kitaev \cite{Kitaev97,AharonovK+98}.)
As a consequence, the supremum in the definition of the diamond norm can be replaced by a maximum.
Moreover, if $\Phi$ is Hermitian-preserving and $\dim(\cW)\geq\dim(\cX)$ then the maximum in the definition of $\tnorm{\Phi\ot\idsup{\cW}}$ is always achieved by some positive semidefinite operator $X$ \cite{RosgenW05,Watrous05,GilchristL+05}.
Thus, if $\Phi$ is Hermitian-preserving then its diamond norm is given by
\begin{align*}
  \dnorm{\Phi}
  &= \max \Tnorm{\Pa{\Phi\ot\idsup{\cW}}(\rho)} \\
  &= \max \Inner{P_0-P_1}{\Pa{\Phi\ot\idsup{\cW}}(\rho)}
\end{align*}
where the maxima in these two expressions are taken over all spaces $\cW$ with dimension at most $\cX$, all states $\rho\in\pos{\cX\ot\cW}$, and all two-outcome measurements $\set{P_0,P_1}\subset\mea{\cY\ot\cW}$.

The diamond norm is to channels as the trace norm is to states:
it provides a physically meaningful distance measure for channels in the sense that the value $\dnorm{\Phi_0-\Phi_1}$ quantifies the observable difference between two channels $\Phi_0,\Phi_1$.
As before, this fact may be illustrated with a simple example.
Suppose Bob selects a channel from $\Phi\in\set{\Phi_0,\Phi_1}$ uniformly at random.
Alice is granted ``one-shot, black-box'' access to $\Phi$ and her goal is to correctly guess which of $\Phi_0,\Phi_1$ was applied.
Specifically, Alice may prepare a quantum system in state $\rho$ and send a portion of that system to Bob, who applies $\Phi$ to that portion and then returns it to Alice.
Finally, Alice performs a two-outcome measurement $\set{P_0,P_1}$ on the resulting state $\Pa{\Phi\ot\identity}(\rho)$ where outcome $a\in\set{0,1}$ indicates a guess that $\Phi=\Phi_a$.

Repeating the derivation from Section \ref{sec:new-norm:trace-norm}, the probability with which Alice successfully discriminates $\Phi_0$ from $\Phi_1$ is seen to be
\[
  \Pr[\textrm{Alice guesses correctly}] =
  \frac{1}{2} + \frac{1}{4}
  \inner{P_0-P_1}{
    \Pa{\Phi_0\ot\identity}(\rho)-
    \Pa{\Phi_1\ot\identity}(\rho)
  } \leq 
  \frac{1}{2} + \frac{1}{4} \dnorm{\Phi_0-\Phi_1}
\]
with equality achieved at the optimal input state $\rho$ and measurement $\set{P_0,P_1}$ for Alice.

It is interesting to note that the ability to send only \emph{part} of the input state $\rho$ to Bob and keep the rest for herself can enhance Alice's ability to distinguish some pairs of channels, as compared to a simpler test that involves sending the \emph{entire} input state to Bob.
Indeed, there exist pairs $\Phi_0,\Phi_1$ of channels that are perfectly distinguishable when applied to half of a maximally entangled input state---that is, $\dnorm{\Phi_0-\Phi_1}=2$---yet they appear nearly identical when an auxiliary system is not used---that is, $\tnorm{\Phi_0-\Phi_1}\approx 0$.
An example of such a pair of linear maps can be found in Watrous \cite{Watrous08}, along with much of the discussion that has occurred thus far in this section.
It is this phenomenon that renders the linear map trace norm less useful than the diamond norm in the study of quantum information.

One might also consider an interpolation between $\tnorm{\Phi}$ and $\dnorm{\Phi}$ in which the dimension of the auxiliary space $\cW$ is restricted to be at most $k$ for some $1\leq k\leq \dim(\cX)$.
Johnston \emph{et al.}\ studied the relationship between these norms and $k$-minimal operator spaces \cite{JohnstonK+11}.
They also showed that for each $k$ the same norm is achieved by replacing the restriction $\dim(\cW)\leq k$ with the restriction that the Schmidt rank of the input state $\rho$ be no larger than $k$.
Timoney \cite{Timoney03} and Watrous \cite{Watrous08} studied conditions on $\Phi$ and $k$ under which this norm is equal to the diamond norm.
(In quantum information theoretic terms, their results esablish conditions under which an auxiliary space $\cW$ of dimension $k$ is sufficient for optimal distinguishability of two channels.)

\subsection{The strategy $r$-norm as a distance measure for strategies}
\label{sec:new-norm:strategy-norm}

The simple guessing game played by Alice and Bob extends naturally from channels to strategies.
Let $S_0,S_1$ be arbitrary $r$-round strategies
and suppose Bob selects $S\in\set{S_0,S_1}$ uniformly at random.
Alice's task is to interact with Bob and then decide after the interaction whether Bob selected $S=S_0$ or $S=S_1$.

Thanks to the inner product relationship between measuring strategies and co-strategies, much of discussion from Section \ref{sec:new-norm:trace-norm} concerning the task of discriminating \emph{states} can be re-applied to the task of discriminating \emph{strategies}.
In particular, Alice can be assumed to act according to some two-outcome $r$-round measuring co-strategy $\set{T_0,T_1}$ for Bob's input and output spaces, with outcome $a\in\set{0,1}$ indicating a guess that Bob acted according to strategy $S_a$.
As before, the probability with which Alice guesses correctly is given by
\[ \Pr[\textrm{Alice guesses correctly}] = \frac{1}{2} + \frac{1}{4} \inner{T_0-T_1}{S_0-S_1}. \]
Naturally, Alice maximizes her chance of success by maximizing this expression over all $r$-round measuring co-strategies $\set{T_0,T_1}$.

Of course, this guessing game is symmetric with respect to strategies and co-strategies.
In particular, if Bob's actions $S_0,S_1$ are \emph{co-strategies} instead of strategies then Alice's actions $\set{T_0,T_1}$ must be a measuring \emph{strategy} instead of a measuring co-strategy.
Alice's maximum success probability is given by the same formula, except that Alice now maximizes this probability over all $r$-round measuring \emph{strategies} $\set{T_0,T_1}$.

With this discrimination problem in mind, the distance measure of Ref.~\cite{ChiribellaD+08b} is recast in the present paper in the form of two norms---one that captures the distinguishability of strategies and one that captures the distinguishability of co-strategies.

\begin{definition}[Strategy $r$-norm---see Ref.~\cite{ChiribellaD+08b}]
\label{def:new-norm}

For any Hermitian operator $X\in\her{\kprod{\cY}{1}{r}\ot\kprod{\cX}{1}{r}}$ let
\begin{align*}
\Snorm{X}{r} \ &\defeq \ \max \Set{ \Inner{T_0-T_1}{X}
: \textrm{$\set{T_0,T_1}$ is an $r$-round measuring co-strategy} },\\
\Snorm{X}{r}^* \ &\defeq \ \max \Set{ \Inner{S_0-S_1}{X}
: \textrm{$\set{S_0,S_1}$ is an $r$-round measuring strategy} }.
\end{align*}
These norms could also be viewed as linear map norms rather than operator norms.
In this case, for any Hermitian-preserving linear map $\Phi:\lin{\kprod{\cX}{1}{r}}\to\lin{\kprod{\cY}{1}{r}}$ one may write
\begin{align*}
  \snorm{\Phi}{r} &\defeq \snorm{\jam{\Phi}}{r},\\
  \snorm{\Phi}{r}^* &\defeq \snorm{\jam{\Phi}}{r}^*.
\end{align*}
The present paper leaves these norms undefined when $X$ is not Hermitian, or, equivalently, when $\Phi$ is not Hermitian-preserving.
\end{definition}

It is not difficult to see that the functions $\snorm{\cdot}{r}$ and $\snorm{\cdot}{r}^*$ are norms.
\begin{proposition}
The functions $\snorm{\cdot}{r}$ and $\snorm{\cdot}{r}^*$ from Definition \ref{def:new-norm} are norms.
\end{proposition}
\begin{proof}
The defining properties of a norm can be verified directly.
It follows immediately from Definition \ref{def:new-norm} that these functions obey the triangle inequality and that they are homogenous (meaning that $\snorm{aX}{r}=\abs{a}\snorm{X}{r}$ for all $a\in\mathbb{R}$).
To see that these functions are positive (meaning that $\snorm{X}{r}\geq 0$ with equality only when $X=0$), it suffices to establish the lower bounds
\begin{align*}
\snorm{X}{r} &\geq \frac{1}{\dim(\kprod{\cX}{1}{r})} \tnorm{X},\\
\snorm{X}{r}^* &\geq \frac{1}{\dim(\kprod{\cY}{1}{r})} \tnorm{X}.
\end{align*}
To this end, let $\Pi_+,\Pi_-$ denote the projections (divided by $\dim(\kprod{\cX}{1}{r})$) onto the positive and nonpositive eigenspaces of $X$, respectively.
Note that
$\Pi_++\Pi_-=\frac{1}{\dim(\kprod{\cX}{1}{r})} I_{\kprod{\cY}{1}{r}\ot\kprod{\cX}{1}{r}},$
which is an $r$-round non-measuring co-strategy.
Hence, $\set{\Pi_+,\Pi_-}$ is an $r$-round measuring co-strategy.
We have
\[ \snorm{X}{r} \geq \Inner{\Pi_+-\Pi_-}{X} = \frac{1}{\dim(\kprod{\cX}{1}{r})} \tnorm{X} \]
as desired.
A similar argument for $\snorm{X}{r}^*$ follows from the observation that $\frac{1}{\dim(\kprod{\cY}{1}{r})} I_{\kprod{\cY}{1}{r}\ot\kprod{\cX}{1}{r}}$ is an $r$-round non-measuring strategy.
\end{proof}

If $S_0,S_1$ are strategies for input spaces $\cX_1,\dots,\cX_r$ and output spaces $\cY_1,\dots,\cY_r$ then it follows immediately from the discussion in Sections \ref{sec:new-norm:trace-norm} and \ref{sec:new-norm:diamond-norm} that the maximum probability with which Alice can correctly distinguish $S_0$ from $S_1$ is
\[ \frac{1}{2} + \frac{1}{4} \snorm{S_0-S_1}{r}. \] 
Likewise, if $S_0,S_1$ are co-strategies rather than strategies then the maximum probability with which Alice can correctly distinguish $S_0$ from $S_1$ is
\[ \frac{1}{2} + \frac{1}{4} \snorm{S_0-S_1}{r}^*. \] 

It may seem superfluous to allow both strategies and co-strategies as descriptions for Bob's actions in this simple example, as every co-strategy may be written as a strategy via suitable relabelling of input and output spaces.
But there is something to be gained by considering both the norms $\snorm{\cdot}{r}$ and $\snorm{\cdot}{r}^*$.
Indeed, it is established by Theorem \ref{thm:unit-ball}
that these norms are dual to each other.

\subsection{Discrimination problems for convex sets of states, channels, and strategies}
\label{sec:new-norm:discrimination-problems}

The guessing game played by Alice and Bob as discussed thus far in this section can be further generalized from a problem of discriminating \emph{individual} states, channels, or strategies to discriminating \emph{convex sets} of states, channels, or strategies.

Specifically, suppose two convex sets $\bA_0,\bA_1$ of states are fixed.
Suppose that Bob arbitrarily selects $\rho_0\in\bA_0$ and $\rho_1\in\bA_1$ and then selects $\rho\in\set{\rho_0,\rho_1}$ uniformly at random and gives Alice a quantum system prepared in state $\rho$.
Alice's goal is to correctly guess whether $\rho\in\bA_0$ or $\rho\in\bA_1$ based upon the outcome of a measurement she conducts on $\rho$.
It is clear that this problem is a generalization of that from Section \ref{sec:new-norm:trace-norm}, as the original problem is recovered by considering singleton sets $\bA_0=\set{\rho_0}$ and $\bA_1=\set{\rho_1}$.

As mentioned in the introduction, this problem of discriminating convex sets of states was solved in Ref.~\cite{GutoskiW05} wherein it was shown that there exists a \emph{single} measurement $\set{P_0,P_1}$ that depends only upon the sets $\bA_0,\bA_1$ with the property that \emph{any} pair $\rho_0\in\bA_0$, $\rho_1\in\bA_1$ may be correctly discriminated with probability at least
\[ \frac{1}{2} + \frac{1}{4} \min_{\sigma_a\in\bA_a} \Tnorm{\sigma_0-\sigma_1}. \]

What can be said about this discrimination problem for convex sets of channels or strategies?
Nothing was known of either problem prior to the work of the present paper.
It is established by Theorem \ref{thm:sep} that the discrimination result for convex sets of states extends unhindered to both channels and strategies.
In particular, it is proven that two convex sets $\bS_0,\bS_1$ of $r$-round strategies can be correctly discriminated with probability at least
\[ \frac{1}{2} + \frac{1}{4} \min_{S_a\in\bS_a} \Snorm{S_0-S_1}{r}.\]
It then follows trivially that two convex sets $\bT_0,\bT_1$ of $r$-round co-strategies can be correctly discriminated with probability at least
\[ \frac{1}{2} + \frac{1}{4} \min_{T_a\in\bT_a} \Snorm{T_0-T_1}{r}^*.\]
As a special case, it holds that two convex sets $\mathbf{\Phi}_0,\mathbf{\Phi}_1$ of channels can be discriminated with probability at least
\[ \frac{1}{2} + \frac{1}{4}\min_{\Phi_a\in\mathbf{\Phi}_a} \Dnorm{\Phi_0-\Phi_1}. \]

\section{Unit ball of the strategy $r$-norm and its dual}
\label{sec:unit-ball-duality}

By employing the characterization of $r$-round strategies mentioned in Section \ref{sec:review}, the quantity $\snorm{X}{r}$ can easily be written as a semidefinite optimization problem:
\begin{align}
\textrm{maximize}\quad & \inner{X}{T_0-T_1} \label{eq:primal}\\
\textrm{subject to}\quad & T_0+T_1 \textrm{ is an $r$-round non-measuring co-strategy} \nonumber \\
&T_0,T_1\succeq 0 \nonumber 
\end{align}
In Appendix \ref{appendix:duality} it is shown that the dual optimization problem is given by
\begin{align}
\textrm{minimize}\quad & p \label{eq:dual}\\
\textrm{subject to}\quad
&{-pS}\preceq X\preceq pS \nonumber\\
& \textrm{$S\succeq 0$ is an $r$-round non-measuring strategy} \nonumber
\end{align}
Moreover, it is also shown in Appendix \ref{appendix:duality} that \emph{strong duality} holds for the optimization problems \eqref{eq:primal}, \eqref{eq:dual}, meaning that these problems have the same optimal value.
Given that, we prove the following theorem.

\begin{theorem}[Unit ball of the strategy $r$-norm and its dual]
\label{thm:unit-ball}

For every Hermitian operator $X\in\her{\kprod{\cY}{1}{r}\ot\kprod{\cX}{1}{r}}$ it holds that
\begin{enumerate}

\item \label{item:unit-ball:st}
$\snorm{X}{r}\leq 1$ if and only if $X=S_0-S_1$ for some $r$-round measuring strategy $\set{S_0,S_1}$.

\item \label{item:unit-ball:cst}
$\snorm{X}{r}^*\leq 1$ if and only if $X=T_0-T_1$ for some $r$-round measuring co-strategy $\set{T_0,T_1}$.

\end{enumerate}
Moreover, the norms $\snorm{\cdot}{r}$ and $\snorm{\cdot}{r}^*$ are dual to each other, meaning that
\begin{align*}
\snorm{X}{r} &= \max_{\snorm{Y}{r}^*\leq 1} \inner{Y}{X}, \\
\snorm{X}{r}^* &= \max_{\snorm{Y}{r}\leq 1} \inner{Y}{X}.
\end{align*}

\end{theorem}

\begin{proof}
We begin with a proof of item \ref{item:unit-ball:st}.
One direction is easy: if $X=S_0-S_1$ for some $r$-round measuring strategy $\set{S_0,S_1}$ then for every $r$-round measuring co-strategy $\set{T_0,T_1}$ it holds that \[ \inner{X}{T_0-T_1} = \inner{S_0-S_1}{T_0-T_1} \leq \inner{S_0+S_1}{T_0+T_1} = 1 \] and so $\snorm{X}{r}\leq 1$.

For the other direction, suppose $\snorm{X}{r}\leq 1$.
By the strong duality of the optimization problems \eqref{eq:primal}, \eqref{eq:dual} (see Appendix \ref{appendix:duality}) there exists an $r$-round non-measuring strategy $S$ with \( -S\preceq X\preceq S. \)
Let \[ S_0 = \frac{1}{2}\Pa{S+X}, \qquad S_1 = \frac{1}{2}\Pa{S-X}. \] 
By construction it holds that $S_0-S_1=X$, that $S_0+S_1=S$, and that $S_0,S_1\succeq 0$.
The proof of item \ref{item:unit-ball:st} is now complete.

That the norm $\snorm{\cdot}{r}^*$ is dual to $\snorm{\cdot}{r}$ now follows immediately:
\[
  \snorm{X}{r}^* =
  \max \Set{ \Inner{S_0-S_1}{X} : \textrm{$\set{S_0,S_1}$ is an $r$-round measuring strategy} }
  = \max_{\snorm{Y}{r}\leq 1} \inner{Y}{X}.
\]
(The first equality is by definition and the second is item \ref{item:unit-ball:st}.)

The remaining claims of the theorem are symmetric to those already proved.
One way to finish the proof would be to formulate a semidefinite optimization problem similar to \eqref{eq:primal} for $\snorm{X}{r}^*$ and then derive its dual as in Appendix \ref{appendix:duality}.
Alternately, the Duality Theorem (see Horn and Johnson \cite{HornJ85}) can be used to achieve a more direct proof.

To that end, note first that the Duality Theorem immediately implies that $\snorm{\cdot}{r}$ is also dual to $\snorm{\cdot}{r}^*$:
\[ \snorm{X}{r} = \max_{\snorm{Y}{r}^*\leq 1} \inner{Y}{X}. \]
To prove item \ref{item:unit-ball:cst}, let $\bB$ denote the set of all operators of the form $T_0-T_1$ for some $r$-round measuring co-strategy $\set{T_0,T_1}$.
We claim that $\bB$ is the unit ball for some norm.
This claim can be established by verifying that the set $\bB$ is compact, convex, symmetric (meaning that $-B\in\bB$ whenever $B\in\bB$), and contains the origin in its interior \cite{HornJ85}.
All but the last of these properties are immediate.
To see that $\bB$ contains the origin in its interior select any Hermitian operator $X$ with
$\norm{X}\leq \frac{1}{\dim(\kprod{\cX}{1}{r})}$ and let
$X=X_+-X_-$ be an orthogonal decomposition of $X$.
Write
\begin{align*}
  D&=\frac{1}{\dim(\kprod{\cX}{1}{r})}I-X_+-X_-,&
  T_0&=X_++\frac{1}{2}D,&
  T_1&=X_-+\frac{1}{2}D.
\end{align*}
Then $\set{T_0,T_1}$ is an $r$-round measuring co-strategy and $X=T_0-T_1$ so $X\in\bB$ and thus $\bB$ contains the origin in its interior.

Let $\norm{\cdot}_\bB$ denote the unique norm whose unit ball is $\bB$.
We already know that
\[ 
  \snorm{X}{r} = \max_{\norm{Y}_\bB\leq 1} \inner{Y}{X} = \max_{\snorm{Y}{r}^*\leq 1} \inner{Y}{X}. \]
(The first equality is by definition and the second by duality of $\snorm{\cdot}{r}$ and $\snorm{\cdot}{r}^*$.)
In particular, each of the norms $\snorm{\cdot}{r}^*$ and $\norm{\cdot}_\bB$ has $\snorm{\cdot}{r}$ as its dual norm.
By the Duality Theorem, these norms must be equal.
\end{proof}

\subsection{Alternate proof of maximum output probabilities}
\label{sec:unit-ball-duality:max-output}

Incidentally, strong duality of the problems \eqref{eq:primal}, \eqref{eq:dual} also yields an alternate proof of a result from Ref.~\cite{GutoskiW07} about maximum output probabilities.
That result is stated as follows.

\begin{theorem}[Maximum output probabilities \cite{GutoskiW07}]
\label{thm:max-output-prob}

Let $\set{S_a}$ be an $r$-round measuring strategy.
The maximum probability with which $\set{S_a}$ can be forced to produce a given outcome $a$ by any $r$-round co-strategy is given by $\snorm{S_a}{r}$.
Furthermore, this quantity equals the minimum value $p$ for which there exists an $r$-round non-measuring strategy $S$ with $S_a\preceq pS$.
An analogous result holds when $\set{S_a}$ is a co-strategy.

\end{theorem}

Theorem \ref{thm:max-output-prob} was originally proven via convex polarity.
While semidefinite optimization duality and convex polarity are nominally different manifestations of the same underlying idea, some readers might be more familiar with semidefinite optimization duality than with convex polarity; the proof presented in the present paper should be more digestible to those readers.

\begin{proof}[New proof of Theorem \ref{thm:max-output-prob}]

It is easy to see that the maximum probability with which $\set{S_a}$ can be forced to produce outcome $a$ is expressed by the semidefinite optimization problem \eqref{eq:primal} with $S_a$ in place of $X$.
(As $S_a\succeq 0$, it is clear that the maximum is attained for operators $T_0,T_1$ with $T_1=0$, implying that $T_0$ is a non-measuring co-strategy.)
By definition, this quantity is $\snorm{S_a}{r}$.

By the strong duality of \eqref{eq:primal}, \eqref{eq:dual}, this quantity equals the minimum over all $p$ such that there exists an $r$-round non-measuring strategy $S$ with $-pS\preceq S_a\preceq pS$.
As $S_a\succeq 0$, the first inequality is trivially satisfied by any $S\succeq 0$ and nonnegative $p$ , and so the theorem follows.
\end{proof}

\section{Distinguishability of convex sets of strategies}
\label{sec:distinguish-convex}

Our proof of the distinguishability of convex sets of strategies is an adaptation of the proof appearing in Ref.~\cite{GutoskiW05}
with states and measurements replaced by strategies and co-strategies and the trace and operator norms replaced with the strategy $r$-norm and its dual.
The requisite properties of these new norms
were established by Theorem \ref{thm:unit-ball}.

\begin{theorem}[Distinguishability of convex sets of strategies]
\label{thm:sep}

Let $\bS_0,\bS_1\subset\pos{\kprod{\cY}{1}{r}\ot\kprod{\cX}{1}{r}}$ be nonempty convex sets of $r$-round strategies.
There exists an $r$-round measuring co-strategy $\set{T_0,T_1}$ with the property that
\[ \Inner{T_0-T_1}{S_0-S_1} \geq \min_{R_a\in\bS_a} \Snorm{R_0-R_1}{r} \]
for all choices of $S_0\in\bS_0$ and $S_1\in\bS_1$.
A similar statement holds in terms of the dual norm $\snorm{\cdot}{r}^*$ for convex sets of co-strategies.

\end{theorem}

\begin{proof}

The proof for co-strategies is completely symmetric to the proof for strategies, so we address only strategies here.
Let $d$ denote the minimum distance between $\bS_0$ and $\bS_1$ as stated in the theorem.
If $d=0$ then the theorem is satisfied by the trivial $r$-round measuring co-strategy corresponding to a random coin flip.
(For this trivial co-strategy, both $T_0$ and $T_1$ are equal to the identity divided by $2\dim(\kprod{\cX}{1}{r})$.)
For the remainder of this proof, we shall restrict our attention to the case $d>0$.

Define
\[ \bS \defeq \bS_0-\bS_1 = \Set{S_0-S_1 : S_0\in\bS_0, S_1\in\bS_1 } \]
and let
\[ \bB \defeq \Set{B\in\her{\kprod{\cY}{1}{r}\ot\kprod{\cX}{1}{r}} : \snorm{B}{r}<d } \]
denote the open ball of radius $d$ with respect to the $\snorm{\cdot}{r}$ norm.
The sets $\bS$ and $\bB$ are nonempty disjoint sets of Hermitian operators, both are convex, and $\bB$ is open.
By the Separation Theorem from convex analysis, there exists a Hermitian operator $H$ and a scalar $\alpha$ such that
\[ \inner{H}{S} \geq \alpha > \inner{H}{B} \]
for all $S\in\bS$ and $B\in\bB$.

For every choice of $B\in\bB$ it holds that $-B\in\bB$, from which it follows that
$\abs{\inner{H}{B}}<\alpha$ for all $B\in\bB$ and hence $\alpha>0$.
Moreover, as $\bB$ is the open ball of radius $d$ in the norm $\snorm{\cdot}{r}$, it follows from the duality of the strategy $r$-norms (Theorem \ref{thm:unit-ball}) that
\[\snorm{H}{r}^* \leq \alpha/d. \]
Now let $\hat{H}=\frac{d}{\alpha}H$ be the normalization of $H$ so that $\snorm{\hat{H}}{r}^*\leq 1$.
It follows from Theorem \ref{thm:unit-ball} that \[ \hat{H}=T_0-T_1 \] for some $r$-round measuring co-strategy $\set{T_0,T_1}$.
It remains only to verify that $\set{T_0,T_1}$ has the desired property:
for every choice of $S_0\in\bS_0$ and $S_1\in\bS_1$ we have
\[ \inner{T_0-T_1}{S_0-S_1} = \inner{\hat{H}}{S_0-S_1} = \frac{d}{\alpha} \inner{H}{S_0-S_1} \geq d \]
as desired.
\end{proof}

The claimed result regarding the distinguishability of convex sets of strategies now follows immediately.
To recap, let $\bS_0,\bS_1$ be convex sets of strategies and let $\set{T_0,T_1}$ denote the measuring co-strategy from Theorem \ref{thm:sep} that distinguishes elements in $\bS_0$ from elements in $\bS_1$.
Suppose Bob selects $S_0\in\bS_0$ and $S_1\in\bS_1$ arbitrarily and then selects $S\in\set{S_0,S_1}$ uniformly at random.
As derived in Section \ref{sec:new-norm}, if Alice acts according to $\set{T_0,T_1}$ then the probability with which she correctly guesses whether $S\in\bS_0$ or $S\in\bS_1$ is given by
\[ \frac{1}{2} + \frac{1}{4}\inner{T_0-T_1}{S_0-S_1} \geq \frac{1}{2} + \frac{1}{4}\min_{R_a\in\bS_a}\Snorm{R_0-R_1}{r}\]
as desired.

\appendix

\section{Appendix to Section \ref{sec:unit-ball-duality}: formal proof of semidefinite optimization duality} \label{appendix:duality}

This appendix contains a formal proof that the semidefinite optimization problems \eqref{eq:primal}, \eqref{eq:dual} from Section \ref{sec:unit-ball-duality} satisfy strong duality.
In other words, their optimal values are equal and are both achieved by feasible solutions.

\subsection{Review of the linear map form for semidefinite optimization}

The semidefinite optimization problem discussed in this appendix is expressed in \emph{linear map form}.
While the linear map form differs superficially from the more conventional \emph{standard form} for these problems, the two forms can be shown to be equivalent and the linear map form is more convenient for our purpose.
Watrous provides a helpful overview of this form of semidefinite optimization \cite{Watrous09}.
For completeness, that overview is reproduced here.

A \emph{semidefinite optimization problem} for spaces $\cP,\cQ$ is specified by a triple $(\Psi,A,B)$ where $\Psi:\lin{\cP}\to\lin{\cQ}$ is a Hermitian-preserving linear map and $A\in\her{\cP}$ and $B\in\her{\cQ}$.
This triple specifies two optimization problems:
\begin{align*}
\textrm{\underline{Primal problem}} && \textrm{\underline{Dual problem}} \\
\textrm{maximize} \quad & \inner{A}{P} & \textrm{minimize} \quad & \inner{B}{Q} \\
\textrm{subject to} \quad & \Psi(P) \preceq B & \textrm{subject to} \quad & \Psi^*(Q) \succeq A\\
& P \in \pos{\cP} & & Q \in \pos{\cQ}
\end{align*}
(Here $\Psi^*:\lin{\cQ}\to\lin{\cD}$ denotes the adjoint of $\Psi$.)
An operator $P$ obeying the constraints of the primal problem is said to be \emph{primal feasible}, while an operator $Q$ obeying the constraints of the dual problem is called \emph{dual feasible}.
The functions $P\mapsto\inner{A}{P}$ and $Q\mapsto\inner{B}{Q}$ are called the primal and dual \emph{objective functions}, respectively.
Let
\begin{align*}
  \alpha &\defeq \sup \Set{ \inner{A}{P}: \textrm{ $P$ is primal feasible} } \\
  \beta  &\defeq \inf \Set{ \inner{B}{Q}: \textrm{ $Q$ is dual feasible} }
\end{align*}
denote the \emph{optimal values} of the primal and dual problems.
(If there are no primal or dual feasible operators then we adopt the convention $\alpha=-\infty$ and $\beta=\infty$, respectively.)

Semidefinite optimization problems derive great utility from the notions of \emph{weak} and \emph{strong duality}.
Weak duality asserts that $\alpha\leq\beta$ for all triples $(\Psi,A,B)$, whereas strong duality provides conditions on $(\Psi,A,B)$ under which $\alpha=\beta$.
Two such conditions are stated explicitly as follows.

\begin{fact}[Strong duality conditions---see Ref.~\cite{BoydV04}]
\label{fact:strong-duality}

  Let $(\Psi,A,B)$ be a semidefinite optimization problem.
  The following hold:
  \begin{enumerate}
  
  \item \label{item:strong-primal}
    (Strict primal feasibility.)
    Suppose $\beta$ is finite and there exists $P\succ 0$ with $\Psi(P)\prec B$.
    Then $\alpha=\beta$ and $\beta$ is achieved by some dual feasible operator.

  \item \label{item:strong-dual}
    (Strict dual feasibility.)
    Suppose $\alpha$ is finite and there exists $Q\succ 0$ with $\Psi^*(Q)\succ A$.
    Then $\alpha=\beta$ and $\alpha$ is achieved by some primal feasible operator.
  
  \end{enumerate}
  
\end{fact}

\subsection{A semidefinite optimization problem for the strategy $r$-norm}
\label{sub-appendix:primal}

Let us construct a triple $(\Psi,A,B)$ whose primal problem is equivalent to the problem \eqref{eq:primal} from Section \ref{sec:unit-ball-duality}.
To this end, it is helpful to observe that \eqref{eq:primal} can be written more explicitly via the linear characterization of co-strategies mentioned in Section \ref{sec:review}:
\begin{alignat*}{2}
\textrm{maximize}\quad & \inner{X}{T_0-T_1}\\
\textrm{subject to}\quad &
T_0+T_1 = T_\s{r}\ot I_{\cY_r} \\
&\Ptr{\cX_r}{T_\s{r}} = T_\s{r-1}\ot I_{\cY_{r-1}} \\
&\qquad\vdots\\
&\Ptr{\cX_2}{T_\s{2}} = T_\s{1}\ot I_{\cY_1} \\
&\Ptr{}{T_\s{1}}=1 \\[2mm]
&T_0,T_1\in\pos{\kprod{\cY}{1}{r}\ot\kprod{\cX}{1}{r}}\\
&T_\s{i}\in\pos{\kprod{\cY}{1}{i-1}\ot\kprod{\cX}{1}{i}} & \qquad (1\leq i\leq r)
\end{alignat*}
The triple $(\Psi,A,B)$ is chosen so that its primal problem captures an inequality relaxation of the above problem.
The components of $(\Psi,A,B)$ are most conveniently expressed in block diagonal form
via the intuitive shorthand notation $\diag(\cdot)$ defined so that, for example,
\[
  \diag\Pa{P,P'} \defeq
  \diag \Pa{
    \begin{array}{c}
      P\\P'
    \end{array}
  } \defeq
  \Pa{
    \begin{array}{cc}
      P&0\\0&P'
    \end{array}
  }.
\]
The operators $A,B$ are given by
\[
  A = \diag\Pa{X,-X,0,\dots,0}
  \qquad
  B = \diag\Pa{0,\dots,0,1}
\]
and the linear map $\Psi$ is given by
\[
  \Psi : \diag\Pa{\begin{array}{c}T_0\\T_1\\T_\s{r}\\\vdots\\T_\s{1}\end{array}}
  \mapsto \diag
  \Pa{
    \begin{array}{c}
      T_0+T_1-T_\s{r}\ot I_{\cY_r} \\
      \Ptr{\cX_r}{T_\s{r}} - T_\s{r-1}\ot I_{\cY_{r-1}} \\
      \vdots \\
      \Ptr{\cX_2}{T_\s{2}} - T_\s{1}\ot I_{\cY_1} \\
      \Ptr{}{T_\s{1}}
    \end{array}
  }
\]

It is straightforward but tedious to verify that the optimal value of the primal problem described by $(\Psi,A,B)$ is equal to $\snorm{X}{r}$.
To this end, let $T$ be any positive semidefinite operator with diagonal blocks $T_0,T_1,T_\s{r},\dots,T_\s{1}$.
The primal objective value at $T$ is given by \[ \inner{A}{T} = \inner{X}{T_0} + \inner{-X}{T_1} = \inner{X}{T_0-T_1} \] as desired, so it remains only to verify that the constraint $\Psi(T)\preceq B$ enforces the property that $T_0+T_1$ is an $r$-round non-measuring co-strategy.
The following lemma serves that purpose.

\begin{lemma}[Correctness of the primal problem] \label{lm:primal}

The optimal value of the primal problem $(\Psi,A,B)$ is achieved by a primal feasible solution $T^\star$ whose diagonal blocks $T_0^\star,T_1^\star,T_\s{r}^\star,\dots,T_\s{1}^\star$ have the property that $T_0^\star+T_1^\star$ is an $r$-round non-measuring co-strategy.

\end{lemma}

\begin{proof}

The proof is a standard ``slackness'' argument: any feasible solution with unsaturated inequality constraints can be ``inflated'' so as to saturate all the constraints without decreasing the objective value of that solution.

Formally, we begin by observing that the optimal value must be achieved by some primal feasible $T$, as the set of feasible solutions is easily seen to be compact.
(In particular, each block of $T$ has trace not exceeding $\dim(\kprod{\cY}{1}{r})$.)
Let $T_0,T_1,T_\s{r},\dots,T_\s{1}$ denote the diagonal blocks of $T$.
As $T$ is primal feasible it holds that $\Psi(T)\preceq B$ and hence
\begin{align*}
  T_0+T_1 &\preceq T_\s{r}\ot I_{\cY_r} \\
  \Ptr{\cX_r}{T_\s{r}} &\preceq T_\s{r-1}\ot I_{\cY_{r-1}} \\
  \vdots \\
  \Ptr{\cX_2}{T_\s{2}} &\preceq T_\s{1}\ot I_{\cY_1} \\
  \Ptr{}{T_\s{1}} &\leq 1.
\end{align*}
To prove the lemma it suffices to construct a feasible solution $T^\star$ whose objective value equals that of $T$ and whose diagonal blocks $T_0^\star,T_1^\star,T_\s{r}^\star,\dots,T_\s{1}^\star$ meet the above constraints with equality.

To this end, the desired blocks $T_\s{1}^\star,\dots,T_\s{r}^\star$ are constructed inductively from $T_\s{1},\dots,T_\s{r}$ so as to satisfy $T_\s{i}^\star\succeq T_\s{i}$ for each $i=1,\dots,r$.
For the base case, it is clear that there is a $T_\s{1}^\star\succeq T_\s{1}$ with $\ptr{}{T_\s{1}^\star}=1$.
For the inductive step, it holds that
\[ \Ptr{\cX_i}{T_\s{i}}\preceq T_\s{i-1}\ot I_{\cY_{i-1}} \preceq T_\s{i-1}^\star\ot I_{\cY_{i-1}}.
\]
(Here we have used the operator inequality
\( P^\star\succeq P \implies P^\star\ot I \succeq P\ot I, \)
an observation that follows from the fact that
\( A\succeq 0 \implies A\ot I\succeq 0 \) by substituting $A=P^\star-P$.)
Thus, there must exist $Q\succeq 0$ with \[ \ptr{\cX_i}{T_\s{i}}+Q=T_\s{i-1}^\star\ot I_{\cY_{i-1}}. \]
Choose any $R\succeq 0$ with $\ptr{\cX_i}{R}=Q$.
Selecting $T_\s{i}^\star=T_\s{i}+R$, it holds that $T_\s{i}^\star\succeq T_\s{i}$ and \[ \Ptr{\cX_i}{T_\s{i}^\star}=T_\s{i-1}^\star\ot I_{\cY_{i-1}} \] as claimed.

The final blocks $T_0^\star,T_1^\star$ are constructed similarly.
As $T_\s{r}^\star\succeq T_\s{r}$, it holds that
\[ T_0+T_1 \preceq T_\s{r}\ot I_{\cY_r} \preceq T_\s{r}^\star\ot I_{\cY_r} \]
and hence there exists $D\succeq 0$ with
\[ T_0+T_1+D=T_\s{r}^\star\ot I_{\cY_r}. \]
Selecting \[ T_0^\star=T_0+\frac{1}{2}D,\qquad T_1^\star=T_1+\frac{1}{2}D, \]
it holds that $T_0^\star,T_1^\star\succeq 0$, that $T_0^\star+T_1^\star$ is an $r$-round co-strategy, and that \( T_0^\star-T_1^\star=T_0-T_1, \) from which it follows that $T^\star$ and $T$ have the same objective value.
\end{proof}

\subsection{The dual problem}

In this section it is shown that the dual problem for $(\Psi,A,B)$ is equivalent to the optimization problem \eqref{eq:dual} from Section \ref{sec:unit-ball-duality}.
To this end, it is helpful to observe that \eqref{eq:dual} can be written more explicitly via the linear characterization of strategies mentioned in Section \ref{sec:review}:
\begin{alignat*}{2}
\textrm{minimize}\quad & p\\
\textrm{subject to}\quad
&S_\s{r}\succeq\pm X \\
&\Ptr{\cY_r}{S_\s{r}} = S_\s{r-1} \ot I_{\cX_r} \\
&\qquad\vdots \\
&\Ptr{\cY_2}{S_\s{2}} = S_\s{1} \ot I_{\cX_2} \\
&\Ptr{\cY_1}{S_\s{1}} = pI_{\cX_1} \\[2mm]
&S_\s{i}\in\pos{\kprod{\cY}{1}{i}\ot\kprod{\cX}{1}{i}} & \qquad (1\leq i\leq r) \\
&p\geq 0
\end{alignat*}

In order to demonstrate the desired equivalence between \eqref{eq:dual} and the dual problem for $(\Psi,A,B)$ we require an explicit formula for the adjoint linear map $\Psi^*$.
It is straightforward but tedious to derive such a formula.
To this end, let $S,T$ be operators with diagonal blocks $S_\s{r},\dots,S_\s{1},p$ and $T_0,T_1,T_\s{r},\dots,T_\s{1}$, respectively.
As $\inner{\Psi(T)}{S}=\inner{T}{\Psi^*(S)}$, a formula for $\Psi^*$ may be derived by writing $\inner{\Psi(T)}{S}$ in terms of the blocks of $T$:
\begin{align*}
  & \Inner{\Psi(T)}{S} \\
  ={}&
    \underbrace{ \Inner{T_0+T_1-T_\s{r}\ot I_{\cY_r}}{S_\s{r}} }_{ \textrm{expand} } +
    \Pa{\sum_{i=1}^{r-1}
      \underbrace{ \Inner{\Ptr{\cX_{i+1}}{T_\s{i+1}} - T_\s{i}\ot I_{\cY_i}}{S_\s{i}} }_{ \textrm{expand} }
    } +
    p\Ptr{}{T_\s{1}} \\
  ={}& \Inner{T_0}{S_\s{r}} + \Inner{T_1}{S_\s{r}} -
    \underbrace{ \Inner{T_\s{r}\ot I_{\cY_r}}{S_\s{r}} }_{ \textrm{isolate $T_\s{r}$} } +
    \Pa{ \sum_{i=1}^{r-1}
      \underbrace{ \Inner{\Ptr{\cX_{i+1}}{T_\s{i+1}}}{S_\s{i}} }_{ \textrm{isolate $T_\s{i+1}$} } -
      \underbrace{ \Inner{T_\s{i}\ot I_{\cY_i}}{S_\s{i}} }_{ \textrm{isolate $T_\s{i}$} }
    } +
    \underbrace{ p\Ptr{}{T_\s{1}} }_{ \textrm{isolate $T_\s{1}$} } \\
  ={}& \Inner{T_0}{S_\s{r}} + \Inner{T_1}{S_\s{r}} -
    \underbrace{ \Inner{T_\s{r}}{\Ptr{\cY_r}{S_\s{r}}} }_{ \substack{\textrm{absorb into summation,} \\ \textrm{remove $T_\s{1}$ term}} } +
    \Pa{\sum_{i=1}^{r-1}
      \Inner{T_\s{i+1}}{S_\s{i}\ot I_{\cX_{i+1}}} - \Inner{T_\s{i}}{\Ptr{\cY_i}{S_\s{i}}}
    } +
    \Inner{T_\s{1}}{pI_{\cX_1}} \\
  ={}&
    \Inner{T_0}{S_\s{r}} + \Inner{T_1}{S_\s{r}} +
    \Pa{\sum_{i=2}^r
      \underbrace{ \Inner{T_\s{i}}{S_\s{i-1}\ot I_{\cX_i}} - \Inner{T_\s{i}}{\Ptr{\cY_i}{S_\s{i}}} }_{ \textrm{collect $T_\s{i}$ terms} }
    } +
    \underbrace{ \Inner{T_\s{1}}{pI_{\cX_1}} - \Inner{T_\s{1}}{\Ptr{\cY_1}{S_\s{1}}} }_{ \textrm{collect $T_\s{1}$ terms} } \\
  ={}&
    \Inner{T_0}{S_\s{r}} + \Inner{T_1}{S_\s{r}} +
    \Pa{\sum_{i=2}^r
      \Inner{T_\s{i}}{S_\s{i-1}\ot I_{\cX_i} - \Ptr{\cY_i}{S_\s{i}}}
    } +
    \Inner{T_\s{1}}{pI_{\cX_1}-\Ptr{\cY_1}{S_\s{1}}}.
\end{align*}
It is now clear that $\Psi^*$ is given by
\[
  \Psi^* : \diag\Pa{\begin{array}{c}S_\s{r}\\\vdots\\S_\s{1}\\p\end{array}}
  \mapsto \diag
  \Pa{
    \begin{array}{c}
      S_\s{r} \\
      S_\s{r} \\
      S_\s{r-1}\ot I_{\cX_r} - \Ptr{\cY_r}{S_\s{r}}\\
      \vdots \\
      S_\s{1}\ot I_{\cX_2} - \Ptr{\cY_2}{S_\s{2}}\\
      p I_{\cX_1} - \Ptr{\cY_1}{S_\s{1}}
    \end{array}
  }.
\]

As was done in section \ref{sub-appendix:primal} for the primal problem, it is now argued that the dual problem for $(\Psi,A,B)$ is an inequality relaxation of \eqref{eq:dual}.
To this end, Let $S$ be any positive semidefinite operator with diagonal blocks $S_\s{r},\dots,S_\s{1},p$.
The dual objective value at $S$ is given by
\[ \inner{B}{S} = \inner{1}{p} = p \]
as desired, so it remains only to verify that the constraint $\Psi^*(S)\succeq A$ enforces the property that $S_\s{r}$ is an $r$-round non-measuring strategy multiplied by $p$.
The following lemma serves that purpose.

\begin{lemma}[Correctness of the dual problem] \label{lm:dual}

For each dual feasible solution $S$ to $(\Psi,A,B)$ (including optimal or near-optimal solutions) there exists another dual feasible solution $S^\star$ whose objective value $p^\star$ equals that of $S$ and whose diagonal blocks $S^\star_\s{r},\dots,S^\star_\s{1},p^\star$ have the property that $S^\star_\s{r}$ is an $r$-round non-measuring strategy multiplied by $p^\star$.

\end{lemma}

\begin{proof}
The proof closely follows the slackness argument used in the proof of Lemma \ref{lm:primal}.
Let $S_\s{r},\dots,S_\s{1},p$ denote the diagonal blocks of $S$.
As $S$ is dual feasible it holds that $\Psi^*(S)\succeq A$ and hence
\begin{align*}
  S_\s{r} &\succeq \pm X\\
  S_\s{r-1}\ot I_{\cX_r} &\succeq \ptr{\cY_r}{S_\s{r}}\\
  \vdots \\
  S_\s{1}\ot I_{\cX_2} &\succeq \ptr{\cY_2}{S_\s{2}}\\
  pI_{\cX_1} &\succeq \ptr{\cY_1}{S_\s{1}}.
\end{align*}
To prove the lemma it suffices to construct a dual feasible solution $S^\star$ whose objective value equals that of $S$ and whose diagonal blocks $S_\s{r}^\star,\dots,S_\s{1}^\star,p^\star$ meet the above constraints with equality (except the constraint $S_\s{r}\succeq \pm X$).

To this end, the desired blocks $S_\s{1}^\star,\dots,S_\s{r}^\star$ are constructed inductively from $S_\s{1},\dots,S_\s{r}$ so as to satisfy $S_\s{i}^\star\succeq S_\s{i}$ for each $i=1,\dots,r$.
For the base case, it is clear that there is an $S_\s{1}^\star\succeq S_\s{1}$ with $pI_{\cX_1}=\ptr{\cY_1}{S_\s{1}^\star}$.
For the inductive step, it holds that
\[ S_\s{i-1}^\star\ot I_{\cX_i}\succeq S_\s{i-1}\ot I_{\cX_i}\succeq\ptr{\cY_i}{S_\s{i}}. \]
({Again, we have used the operator inequality
\( P^\star\succeq P \implies P^\star\ot I \succeq P\ot I \)
for any $P\succeq 0$.})
Thus, there must exist $Q\succeq 0$ with
\[ S_\s{i-1}^\star\ot I_{\cX_i} = \ptr{\cY_i}{S_\s{i}} + Q. \]
Choose any $R\succeq 0$ with $\ptr{\cY_i}{Q}=R$.
Selecting $S_\s{i}^\star = S_\s{i}+R$, it holds that $S_\s{i}^\star\succeq S_\s{i}$ and
\[ S_\s{i-1}^\star\ot I_{\cX_i} = \ptr{\cY_i}{S_\s{i}^\star} \]
as claimed.

Selecting $p^\star=p$, it holds that $S^\star_\s{r}$ is an $r$-round non-measuring strategy multiplied by $p^\star$ as desired.
As $S_\s{r}^\star\succeq S_\s{r}\succeq \pm X$ and $p^\star=p$, it follows that $S^\star$ is a dual feasible solution that achieves the same objective value as $S$.
\end{proof}

\subsection{Strong duality}
Thus far, it has been argued that the optimal values of the problems \eqref{eq:primal}, \eqref{eq:dual} from Section \ref{sec:unit-ball-duality} are captured by the primal and dual semidefinite optimization problems associated with the triple $(\Psi,A,B)$.
It remains only to show that these two quantities are equal.
Equality is established by showing that $(\Psi,A,B)$ satisfies the conditions for strong duality from Fact \ref{fact:strong-duality}.

\begin{theorem}[Strong duality of $(\Psi,A,B)$]
\label{lm:strong-duality}

There exists a primal feasible operator $T$ and a dual feasible operator $S$ such that $\inner{A}{T}=\inner{B}{S}$.

\end{theorem}

\begin{proof}

The proof is via item \ref{item:strong-primal} of Fact \ref{fact:strong-duality} (Strong duality conditions).
Specifically, it is shown that $\beta$ is finite and the primal problem is strictly feasible.
It then follows from Fact \ref{fact:strong-duality} that $\alpha=\beta$ and that $\beta$ is achieved for some dual feasible operator.
To complete the proof, it suffices to note that the optimal value $\alpha$ is also achieved by a primal feasible operator, as established in Lemma \ref{lm:primal} (Correctness of the primal problem).

First, it is argued that $\beta$ is finite.
As $B\succeq 0$, any dual feasible solution has nonnegative objective value.
Thus, to show that $\beta$ is finite it suffices to exhibit a single dual feasible solution.
That solution $S$ is a block-diagonal matrix with blocks $S_\s{r},\dots,S_\s{1},p$ given by
\begin{align*}
S_\s{r} &= \norm{X} I_{\kprod{\cY}{1}{r}\ot\kprod{\cX}{1}{r}}\\
S_\s{r-1} &= \norm{X} \dim(\cY_r) I_{\kprod{\cY}{1}{r-1}\ot\kprod{\cX}{1}{r-1}}\\
&\qquad\vdots\\
S_\s{1} &= \norm{X} \dim(\kprod{\cY}{2}{r}) I_{\cY_1\ot\cX_1}\\
p &= \norm{X} \dim(\kprod{\cY}{1}{r}).
\end{align*}
As $X$ is Hermitian it holds that
\[ -S_\s{r} = -\norm{X} I_{\kprod{\cY}{1}{r}\ot\kprod{\cX}{1}{r}} \preceq X \preceq \norm{X} I_{\kprod{\cY}{1}{r}\ot\kprod{\cX}{1}{r}} = S_\s{r} \]
and hence $S$ is dual feasible as desired.

Finally, it is shown that the primal is strictly feasible.
Choose $\delta\in(0,\frac{1}{r+1})$ and let $T$ be the block-diagonal operator with diagonal blocks $T_0,T_1,T_\s{r},\dots,T_\s{1}$ given by
\begin{alignat*}{2}
  T_\s{i} &= \frac{1-i\delta}{\dim(\kprod{\cX}{1}{i})}I_{\kprod{\cY}{1}{i-1}\ot\kprod{\cX}{1}{i}}
  &\qquad (1\leq i\leq r) \\
  T_0=T_1 &= \frac{1-(r+1)\delta}{2\dim(\kprod{\cX}{1}{r})} I_{\kprod{\cY}{1}{r}\ot\kprod{\cX}{1}{r}}.
\end{alignat*}
It is clear that $T\succ 0$ and it is tedious but straightforward to verify that $\Psi(T)\prec B$.
Specifically, we have
\begin{align*}
  \Ptr{}{T_\s{1}} = 1-\delta &\ < \ 1 \\
  \Ptr{\cX_2}{T_\s{2}} = \frac{1-2\delta}{\dim(\cX_1)}I_{\cY_1\ot\cX_1}
  &\ \prec \ \frac{1-\delta}{\dim(\cX_1)} I_{\cY_1\ot\cX_1} = T_\s{1}\ot I_{\cY_1} \\
  &\ \vdots \\
  \Ptr{\cX_r}{T_\s{r}} = \frac{1-r\delta}{\dim(\kprod{\cX}{1}{r-1})} I_{\kprod{\cY}{1}{r-1}\ot\kprod{\cX}{1}{r-1}}
  &\ \prec \ \frac{1-(r-1)\delta}{\dim(\kprod{\cX}{1}{r-1})} I_{\kprod{\cY}{1}{r-1}\ot\kprod{\cX}{1}{r-1}} = T_\s{r-1}\ot I_{\cY_{r-1}} \\
  T_0 + T_1 = \frac{1-(r+1)\delta}{\dim(\kprod{\cX}{1}{r})} 
  I_{\kprod{\cY}{1}{r}\ot\kprod{\cX}{1}{r}}
  &\ \prec \ \frac{1-r\delta}{\dim(\kprod{\cX}{1}{r})}
  I_{\kprod{\cY}{1}{r}\ot\kprod{\cX}{1}{r}} = T_\s{r}\ot I_{\cY_r}.
\end{align*}
It now follows from item \ref{item:strong-primal} of Fact \ref{fact:strong-duality} that $\alpha=\beta$ and that $\beta$ is achieved by some dual feasible operator.
\end{proof}

\section*{Acknowledgements}

The author is grateful to Giulio Chiribella and John Watrous for informative discussions.
This research was supported by the Government of Canada through Industry Canada, the Province of Ontario through the Ministry of Research and Innovation, NSERC, DTO-ARO, CIFAR, and QuantumWorks.
Part of this research was conducted while the author was a graduate student at the University of Waterloo, at which time this research was supported by Canada's NSERC and the David R.~Cheriton School of Computer Science at the University of Waterloo.


\newcommand{\etalchar}[1]{$^{#1}$}

\end{document}